\documentclass[twocolumn,showpacs,amsmath,prb,amssymb]{revtex4}
\usepackage{graphicx}
\usepackage{latexsym}
\usepackage{subfigure}
\usepackage{tabularx}
\begin{document}

\title{First principles study of the electronic and magnetic structures of the tetragonal and orthorhombic phases of Ca$_{3}$Mn$_{2}$O$_{7}$} 
\author{S. F. Matar\footnote {email: matar@icmcb-bordeaux.cnrs.fr. Web: http://www.m3pec.u-bordeaux1.fr/matar}, V. Eyert $^{(1)}$, A. Villesuzanne, M.-H. Whangbo $^{(2)}$}
\address{Institut de Chimie de la Mati\`{e}re Condens\'{e}e de Bordeaux,
         ICMCB-CNRS, Universit\'e Bordeaux 1, 33608 Pessac Cedex, France}
\address{(1) Institut f\"ur Physik, Universit\"at Augsburg, 86135 Augsburg, 
         Germany}
\address{(2) Department of Chemistry, North Carolina State University, 
         Raleigh, USA}

\date{\today}

\begin{abstract}
On the basis of density functional theory electronic band structure calculations using the augmented spherical wave method, the electronic and magnetic properties of the orthorhombic and tetragonal phases of Ca$_{3}$Mn$_{2}$O$_{7}$ were investigated and the spin exchange interactions of the orthorhombic phase were analyzed. Our calculations show that the magnetic insulating states are more stable than the non-magnetic metallic state for both polymorphs of Ca$_{3}$Mn$_{2}$O$_{7}$, the orthorhombic phase is more stable than the tetragonal phase, and the ground state of the orthorhombic phase is antiferromagnetic. The total energies calculated for the three spin states of the orthorhombic phase of Ca$_{3}$Mn$_{2}$O$_{7}$ led to estimates of the spin exchange interactions J$_{nn}$ = -3.36 meV and J$_{nnn}$ = -0.06 meV. The accuracy of these estimates were tested by calculating the Curie-Weiss temperature within the mean-field approximation. 
\end{abstract}

\maketitle

\section{Introduction} 
Due to their potential technological application, magnetic manganese oxides have received much attention in the past few years [1-3]. It is an important theoretical issue to understand their electronic and magnetic properties [4, 5]. In accounting for such properties of magnetic oxides, density functional theory (DFT) [6, 7] electronic structure calculations using the local spin density approximation (LSDA) [8] and the generalized gradient approximation (GGA) [9] have proven very reliable. In the Ruddlesden-Popper (RP) family of manganites Ca$_{n+1}$Mn$_{n}$O$_{3n+1}$ [10], the diamagnetic CaO rocksalt layers alternate with the magnetic (CaMnO$_{3}$)$_{n}$ layers along the crystallographic c-direction, where n represents the number of MnO$_{3}$ perovskite sheets in each (CaMnO$_{3}$)$_{n}$ layer. To emphasize this structural feature, the formulas for these oxides can be rewritten as (CaO)(CaMnO$_{3}$)$_{n}$ with n = 1, 2, 3, ...,$\infty $  . Known examples of this RP family are Ca$_{2}$MnO$_{4}$ (n = 1), Ca$_{3}$Mn$_{2}$O$_{7}$ (n = 2) and Ca$_{4}$Mn$_{3}$O$_{10}$ (n = 3) [11]. In a recent electronic  structure study [12], it was shown that the n = 1 RP member, Ca$_{2}$MnO$_{4}$, and its reduced phase Ca$_{2}$MnO$_{3.5}$ are antiferromagnetic (AF) insulators with band gap of approximately $\sim$ 1 eV. According to a recent neutron diffraction study [13] at room temperature, the n = 2 RP member, Ca$_{3}$Mn$_{2}$O$_{7}$, exists as a mixture of the majority orthorhombic and the minority tetragonal phases. The main structural difference between the two is depicted in Figure 1. Due to the tilting of the MnO$_{6}$ octahedra, the $\widehat {Mn-O-Mn}$ angles between two adjacent -corner sharing- octahedra are considerably smaller 180$^{o}$ (i.e., 159.1, 163.2 and 163.9$^{o}$). Ca$_{3}$Mn$_{2}$O$_{7}$ undergoes a three-dimensional (3D) magnetic ordering below T$_{N}$ $\sim$  110 - 115 K with a G-type AF structure [13]. As a part of our continuing studies on calcium manganese oxides [4, 12, 14, 15], we examine in the present work the electronic and magnetic structures of the two polymorphs of Ca$_{3}$Mn$_{2}$O$_{7}$ on the basis of first principles DFT electronic  structure calculations. 

\section{Calculations}
Our DFT calculations employed the all-electron augmented spherical wave (ASW) method in its scalar-relativistic implementation[16, 17].  As preliminary calculations revealed that the GGA describes the effects of exchange and correlation better than the LSDA, hence leading to a more accurate description of the energy differences between different ordered magnetic states, we opted for a use of the GGA in the form proposed by Perdew, Burke, and Ernzerhof [9].  In the ASW method, the wave function is expanded in atom-centered augmented spherical waves, which are Hankel functions and numerical solutions of Schr\"odinger's equation, respectively, outside and inside the so-called augmentation spheres. In order to optimize the basis set, additional augmented spherical waves were placed at carefully selected interstitial sites. The choice of these sites as well as the augmentation radii were automatically determined using the sphere-geometry optimization (SGO) algorithm [18]. The Ca 4s, Mn 4s, Mn 3d and O 2p states were treated as valence states, and the low-lying O 2s states as core states. The basis set was complemented by wave functions of s-, p-, and possibly, d-symmetry centered at the interstitial sites. A sufficiently large number of k points was used to sample the irreducible wedge of the Brillouin zone, and the energy and charge differences between successive iterations were converged below  $\Delta$E = 10$^{-8}$ Ryd. and  $\Delta$ Q = 10$^{-8}$, respectively, to obtain accurate values of the magnetic moments and accurate total energy differences between various ordered magnetic states. 
To extract more information about the nature of the interactions between the atomic constituents from electronic  structure calculations, the crystal orbital overlap population (COOP) [19] or the crystal orbital Hamiltonian population (COHP) [20] may be employed. Both approaches provide a qualitative description of the bonding, nonbonding and antibonding interactions between two atoms. A slight refinement of the COHP was recently proposed in form of the "energy of covalent bond" (ECOV), which combines COHP and COOP to calculate quantities independent of the choice of the zero of potential [21]. Both COOP and ECOV give similar general trends, but COOP, when defined within plane-wave basis sets, exaggerates the magnitude of antibonding states. In the present work the ECOV was used for the chemical bonding analysis. In the plots, negative, positive and zero ECOV magnitudes are relevant to bonding, antibonding and nonbonding interactions respectively.

\section{Results and discussion }

\subsection{Non-magnetic configuration}
To gain insight into the chemical bonding in the two forms of Ca$_{3}$Mn$_{2}$O$_{7}$, their non-magnetic (NM) states were calculated by performing non-spin-polarized electronic structure calculations by enforcing spin degeneracy for all species. Our calculations show charge transfer from the cation sites (Ca$^{2+}$, Mn$^{4+}$) to the anionic sites of oxygen  as well as toward the empty space, but the amount of charge transfer is not significant of an ionic character of the atomic constituents. The plots of the projected density of states (DOS) for the constituent atom sites are presented in Figure 2. The projected DOS plots calculated for the five Mn 3d orbitals are presented in Figure 3, and the ECOV plots calculated for the various Mn-O bonds in Figure 4.
 
The comparison of Figures 2 and 4 show that the valence bands (VB's) are dominated by Mn-O bonding. The Mn-O interactions are bonding in the major part of the VB's with strong contributions in the -8, -6 eV region (Figure 2), i.e., where the Mn e$_{g}$-orbitals (Figure 3) make  $\sigma$-bonding with the O 2p orbitals. The s-like Ca states are smeared into the whole range of the VB's (Figure 2). For both polymorphs of Ca$_{3}$Mn$_{2}$O$_{7}$, the Fermi level E$_{F}$ occurs near the DOS peak dominated by the Mn 3d states so that the DOS value at the Fermi level, n(E$_{F}$), is large. The DOS peak around E$_{F}$ arises from the t$_{2g}$-block bands in which the Mn t$_{2g}$-orbitals make weak  $\pi$-antibonding interactions with the O 2p orbitals. The existence of Mn-O antibonding in this DOS peak is clearly seen from the ECOV plot (Figure 4). The occurrence of the Fermi level around a sharp DOS peak with a large n(E$_{F}$) value, indicates an instability towards a ferromagnetic state. This  can be inferred from the Stoner's mean field theory analysis (see [22] for a review on magnetic oxides) whereby a nonzero magnetic moment on the Mn sites should occur when two spin populations are accounted for in the calculations.

\subsection{Magnetic states}
Spin-polarized electronic structure calculations were carried out for the ferromagnetic (FM) state of the orthorhombic and tetragonal phases of Ca$_{3}$Mn$_{2}$O$_{7}$, and for the AF state of the orthorhombic phase of Ca$_{3}$Mn$_{2}$O$_{7}$. There are a number of possible AF spin arrangements. In the present study, we considered two types of AF arrangements, i.e., the G- and A-type AF states shown in Figure 5. In the G-type AF state the antiferromagnetically ordered planes of Mn sites are antiferromagnetically coupled. In the A-type AF arrangement, the ferromagnetically ordered planes of Mn sites are antiferromagnetically coupled within each perovskite layer. 

Results of our calculations for the FM state of the tetragonal and orthorhombic phases of Ca$_{3}$Mn$_{2}$O$_{7}$ are summarized in Table I as well as Figure 6. For both phases the FM state is substantially more stable than the NM states (i.e., E$_{FM}$ $<$ E$_{NM}$), and the total moment per formula unit (FU) is 6.0  $\mu_B$ (Table I), as expected for the high-spin Mn$^{4+}$ (d$^{3}$) ions inCa$_{3}$Mn$_{2}$O$_{7}$. The total energy obtained for the orthorhombic phase is by 170 meV/FU lower than that of  the tetragonal phase, which is consistent with the finding that the orthorhombic and tetragonal phases are the majority and minority phases, respectively. The projected DOS plots calculated for the FM state of the orthorhombic Ca$_{3}$Mn$_{2}$O$_{7}$ are presented in Figure 6a. The up-spin and down-spin bands are strongly split such that the up-spin t$_{2g}$-block bands are completely filled while the down-spin t$_{2g}$-block bands are empty, as expected for the high-spin Mn$^{4+}$ (d$^{3}$) ions. The down- and up- spin bands have a band gap of about 1.0 and 0.3 eV, respectively. Given the fact that both the LSDA and the GGA underestimate band gaps, the FM state of Ca$_{3}$Mn$_{2}$O$_{7}$ should be insulating experimentally. However, the FM state is not the magnetic ground state of Ca$_{3}$Mn$_{2}$O$_{7}$, as will be discussed in next section.  Figure 6b shows the ECOV plots for the Mn-O bonds calculated for the up-spin and down-spin bands of the orthorhombic phase. The VB's exhibit Mn-O bonding interactions up to -2 eV in both up-spin and down-spin bands. The energy region beteween -2 eV and the Fermi level shows Mn-O antibonding interactions due to the $\pi$-type antibonding interactions of the t$_{2g}$-block bands in the up-spin bands, but this feature is missing in the down-spin bands because the down-spin t$_{2g}$-block bands are raised above the Fermi level. Thus, the exchange splitting in the FM state enhances Mn-O bonding interactions in Ca$_{3}$Mn$_{2}$O$_{7}$. 

Results of calculations for the G- and A-type AF statements of the orthorhombic phase of Ca$_{3}$Mn$_{2}$O$_{7}$ are summarized in Table II and Figure 7. Both AF states are more stable than the FM state, and the G-type AF state is slightly more stable than the A-type AF state (Table II). The latter is consistent with the neutron diffraction study [13], which found a G-type AF ordering for Ca$_{3}$Mn$_{2}$O$_{7}$. While the resulting magnetization is zero as expected in AF order, the atomic magnetic moments and the resulting $\uparrow$,$\downarrow$ total magnetization are reduced with respect to their value in the ferromagnetic configuration. This is due to the symmetry breaking when the AF lattice was accounted for. The DOS and band structure are shown in Figure 7 for  one magnetic sublattice. The similarities with the  features observed in Fig. 6 for the ferromagnetic case are present here too. So at least from the point of view of the electronic band structure some relationship is persistent although the AF state is the ground state with an insulating character. From the band structure plot (Figure 7b) the large dispersion of the bands within the CB illustrates furthermore the itinerant character of the e$_{g}$ states. The gap of $\sim$ 0.4 eV is found direct, between $\Gamma$$_{V}$ and $\Gamma$$_{C}$. Since both the LSDA and the GGA underestimate the band gap, the system can be expected to be insulating experimentally.

\section{Analysis of the spin exchange interactions} 
It is of interest to estimate the nearest-neighbor (NN) spin exchange J$_{nn}$ and the next-nearest-neighbor (NNN) spin exchange J$_{nnn}$ interactions of the (CaMnO$_{3}$)$_{2}$ double-perovskite layer in the orthorhombic phase of Ca$_{3}$Mn$_{2}$O$_{7}$. The spin exchange interactions between the adjacent double-perovskite layers are expected to be negligible, so we consider only those interactions within a double-perovskite layer. Then, each Mn site has five NN and eight NNN spin exchange interactions. For simplicity, we assume that all five NN interactions are identical, and so are all the eight NNN interactions. Then, in terms of J$_{nn}$ and J$_{nnn}$, the energies per Mn site of the FM, G-type AF and A-type-AF states are written as [23-25]

\begin{eqnarray}
\nonumber
E(FM) = {\frac{1}{2}}~ (-5J_{nn} - 8J_{nnn})(\frac{N^{2}}{4}) = (\frac{N^{2}}{4})(-2.5J_{nn} - 4J_{nnn}),\\
\nonumber
E(G-AF) = {\frac{1}{2}}~ (5J_{nn} - 8J_{nnn})(\frac{N^{2}}{4}) = (\frac{N^{2}}{4})(2.5J_{nn} - 4J_{nnn}),\\
E(A-AF) = {\frac{1}{2}}~ (-3J_{nn})(\frac{N^{2}}{4}) = (\frac{N^{2}}{4})(-1.5J_{nn}),			
\end{eqnarray}
where N is the number of unpaired spins at each Mn$^{4+}$ site (i.e., N = 3). According to the results of our electronic band structure calculations (Table II), we have

\begin{eqnarray}
\nonumber
	E(FM) - E(G-AF) = 37.8 meV/Mn,\\
	E(A-AF) - E(G-AF) = 29.7 meV/Mn.				
\end{eqnarray}
From Eqs. (1) and (2), it is estimated that J$_{nn}$ = -3.36 meV and J$_{nnn}$ = -0.06 meV, i.e., $\frac{J_{nn}}{k_{B}}$ = -39 K and $\frac{J_{nnn}}{k_{B}}$ = -0.7 K, where k$_{B}$ is the Boltzmann constant. The NN interaction is strongly AF, and the NNN interaction is very weakly AF. Therefore, within the mean-field approximation, the Curie-Weiss temperature   of Ca$_{3}$Mn$_{2}$O$_{7}$ can be estimated from the J$_{nn}$ value according to the expression (3)
\begin{equation}
\Theta  =  \frac{zJ_{nn}S(S+1)}{k_{B}},
\end{equation} 						
where z = 5, and S = $\frac{3}{2}$ for the high-spin Mn$^{3+}$. For J$_{nn}$/k$_{B}$ $\sim$  -39 K, we obtain $\Theta$=    -244 K, which is somewhat smaller in magnitude than the experimental value  $\Theta$$_{exp}$ = - 465 K [26]. This suggests that our calculations underestimated the spin exchange interactions of Ca$_{3}$Mn$_{2}$O$_{7}$ by a factor of approximately two. However such a magnitude is expected in the framework of the approximation used within the DFT and its functionals when comparisons with mean-field analysis results are carried out.  

\section{Concluding remarks and outlook}
Our electronic  structure study reveals that for both polymorphs of Ca$_{3}$Mn$_{2}$O$_{7}$, the magnetic states are more stable than the non-magnetic states. The spin polarization makes Ca$_{3}$Mn$_{2}$O$_{7}$ insulating and lowers its energy mainly through exchange leading to a magnetic moment on Mn expected for a high-spin Mn$^{4+}$ (d$^{3}$) ion. The down-spin states provide more Mn-O bonding than do the up-spin states because the down-spin Mn t$_{2g}$-block bands are raised above the Fermi level. In agreement with experiment, our calculations show that the orthorhombic phase is more stable than the tetragonal phase, and the ground state of the orthorhombic phase is G-type AF. The total energies of the FM, A-type AF and G-type AF states of Ca$_{3}$Mn$_{2}$O$_{7}$ obtained from our calculations lead to the estimates of the spin exchange interactions $\frac{J_{nn}}{k_{B}}$ = -39 K and $\frac{J_{nnn}}{k_{B}}$ = -0.7 K. The Curie-Weiss temperature  $\Theta$ estimated from these spin exchange parameters is approximately half the experimental value, so that our calculations underestimated the values of the spin exchange parameters by a factor of approximately two.

\section{Acknowledgements }
We acknowledge the computational facilities provided by the computer center of the University Bordeaux 1 within the $M3PEC$ (http://www.m3pec.u-bordeaux1.fr) intensive calculations project partly financed by the Conseil R\'egional d'Aquitaine.  M.-H. W. thanks the support by the Office of Basic Energy Sciences, Division of Materials Sciences, U. S. Department of Energy, under Grant DE-FG02-86ER45259. V.E. acknowledges support by the Deutsche Forschungsgemeinschaft through SFB 484.

{}

\newpage

\begin{table}
\caption{Total energies and magnetic moments of the atomic sites calculated for the ferromagnetic states of the tetragonal and orthorhombic phases of Ca$_{3}$Mn$_{2}$O$_{7}$. The energies are given in units of eV per half the FU (i.e., per Mn) with respect to those of the non-spin-polarized state. The magnetic moments are given in units of Bohr magneton ($\mu_B$).}
\bigskip
\begin{center}
\begin{tabular}{lclclcl}
Parameters&		Tetragonal&			Orthorhombic\\
\hline
$\Delta$E$_{FM-NM}$ (eV)& 	- 1.442& 			-1.610	\\
$\Delta$E$_{FM-NM}$ (eV)/Mn& 	- 0.721& 			-0.831	\\
$\mu$(Mn)& 	     2.729& 		2.719\\
$\mu$(O1)& 	     0.088& 		0.092\\
$\mu$(O2)& 	    -0.001& 		0.047\\
$\mu$(O3)& 	     0.073& 		0.056\\
$\mu$(O4)& 	 	 & 		0.014\\
$\mu$(FU)& 	        6.0& 		6.0\\
\hline
\end{tabular}
\end{center}
\label{tab1}
\end{table}

\begin{table}
\caption{Total energies and magnetic moments of the atomic sites calculated for the A- and G-type antiferromagnetic states of the orthorhombic phases of Ca$_{3}$Mn$_{2}$O$_{7}$. The energies are given in units of eV per half the FU (i.e., per Mn) with respect to those of the FM state. The magnetic moments are given in units of Bohr magneton ($\mu_B$)}
\bigskip
\begin{center}
\begin{tabular}{lclclcl}
Parameters&		A-type AF&			G-type AF\\
\hline
$\Delta$E$_{AF-FM}$ (eV)& 	- 0.0082& 			-0.0378	\\
$\mu$(Mn) & 	 $\pm$2.613& 		$\pm$2.678\\
$\mu$(O1) & 	 $\pm$0.058& 		$\sim$~0.0\\
$\mu$(O2)& 	    0.0& 		$\pm$0.03\\
$\mu$(O3)& 	     0.0& 		$\pm$0.05\\
$\mu$(O4)& 	 	0.0 & 		0.02\\
$\mu$(FU)& 	        $\pm$5.168& 	$\pm$5.380\\
\hline
\end{tabular}
\end{center}
\label{tab1}
\end{table}
\begin{figure}[htp]
\centering
\subfigure{\includegraphics[width=0.25\textwidth]{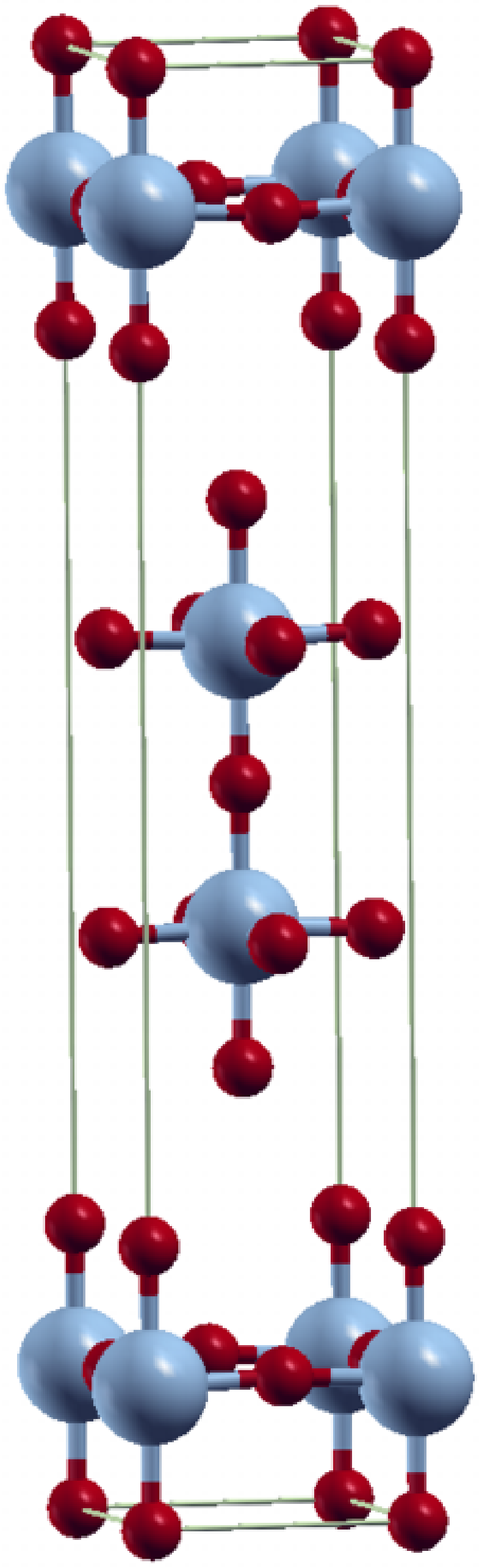}}{a)}
\hspace{0.01\textwidth}
\subfigure{\includegraphics[width=0.35\textwidth]{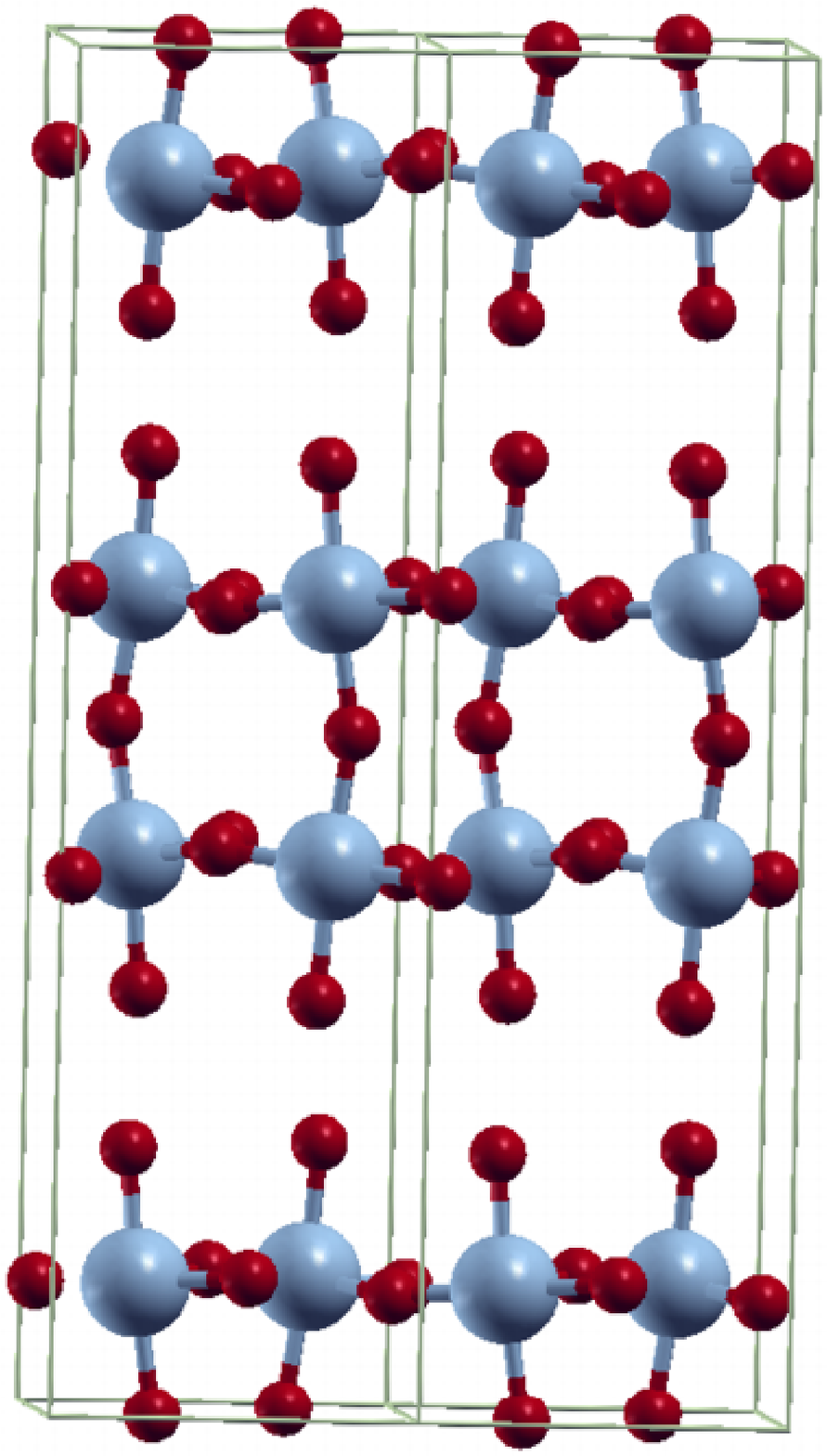}}{b)}
\caption{{\it (color online)} 
        Structures of (a) the minority tetragonal and (b) the majority orthorhombic phases of Ca$_{3}$Mn$_{2}$O$_{7}$.}
\label{fig:struct}
\end{figure}

\begin{figure}[htp]
\centering
\subfigure{\includegraphics[width=0.42\textwidth]{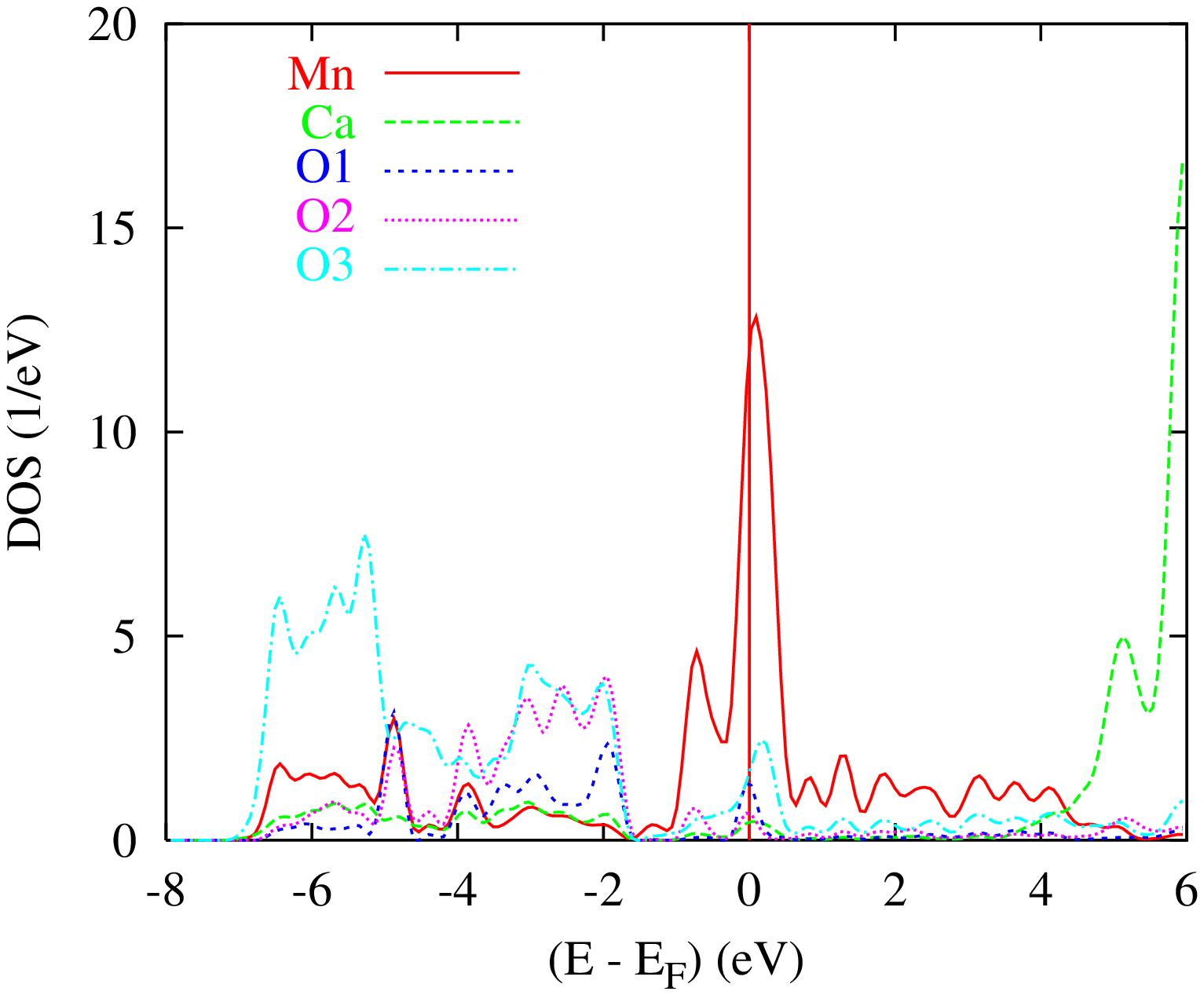}}{a)}
\hspace{0.01\textwidth}
\subfigure{\includegraphics[width=0.42\textwidth]{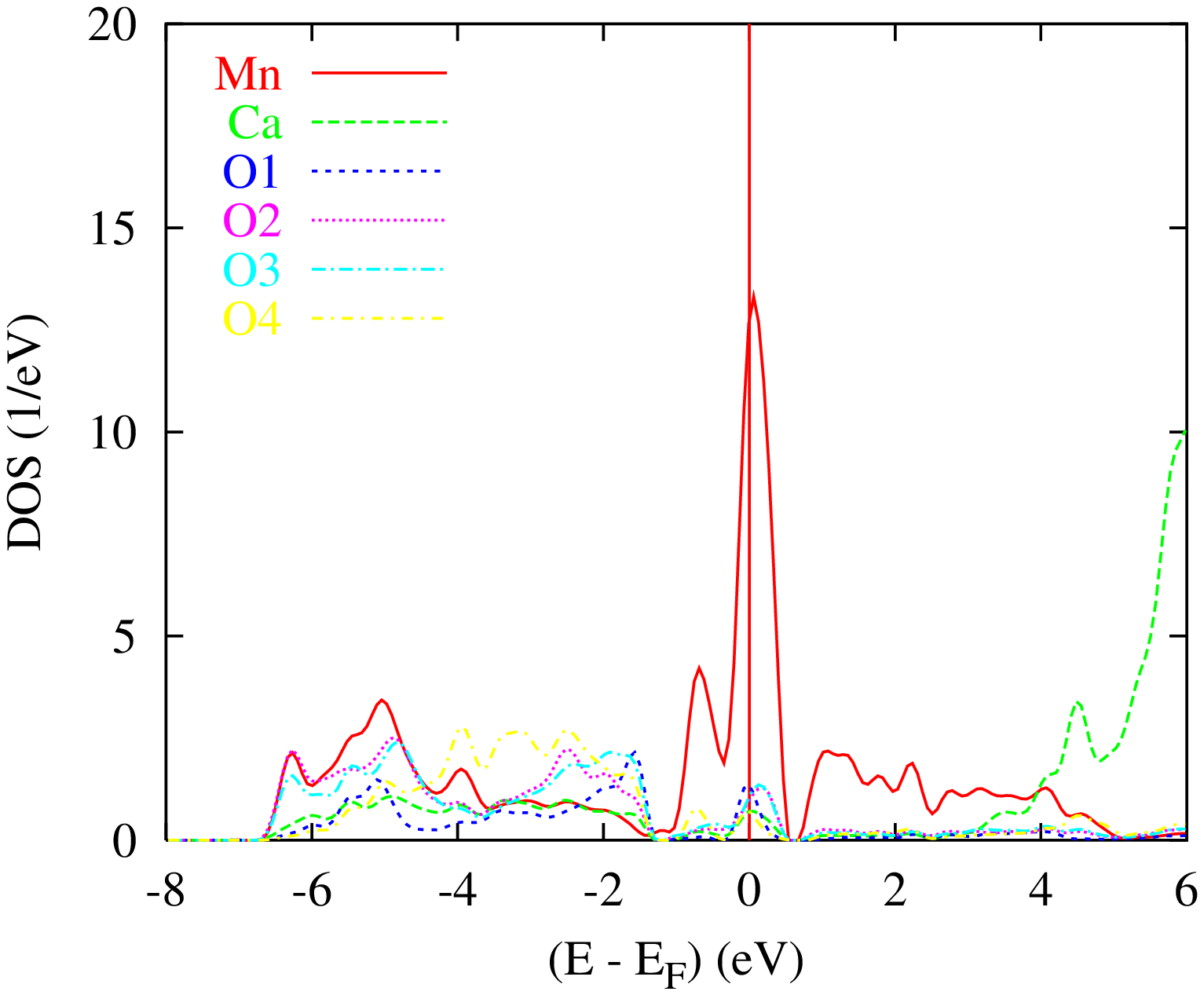}}{b)}
\caption{{\it (color online)}
         Projected DOS plots of the atomic sites calculated for the NM states of (a) the minority tetragonal and (b) the majority orthorhombic phases of Ca$_{3}$Mn$_{2}$O$_{7}$.}
\label{fig:dos}
\end{figure}

\begin{figure}[htp]
\centering
\subfigure{\includegraphics[width=0.42\textwidth]{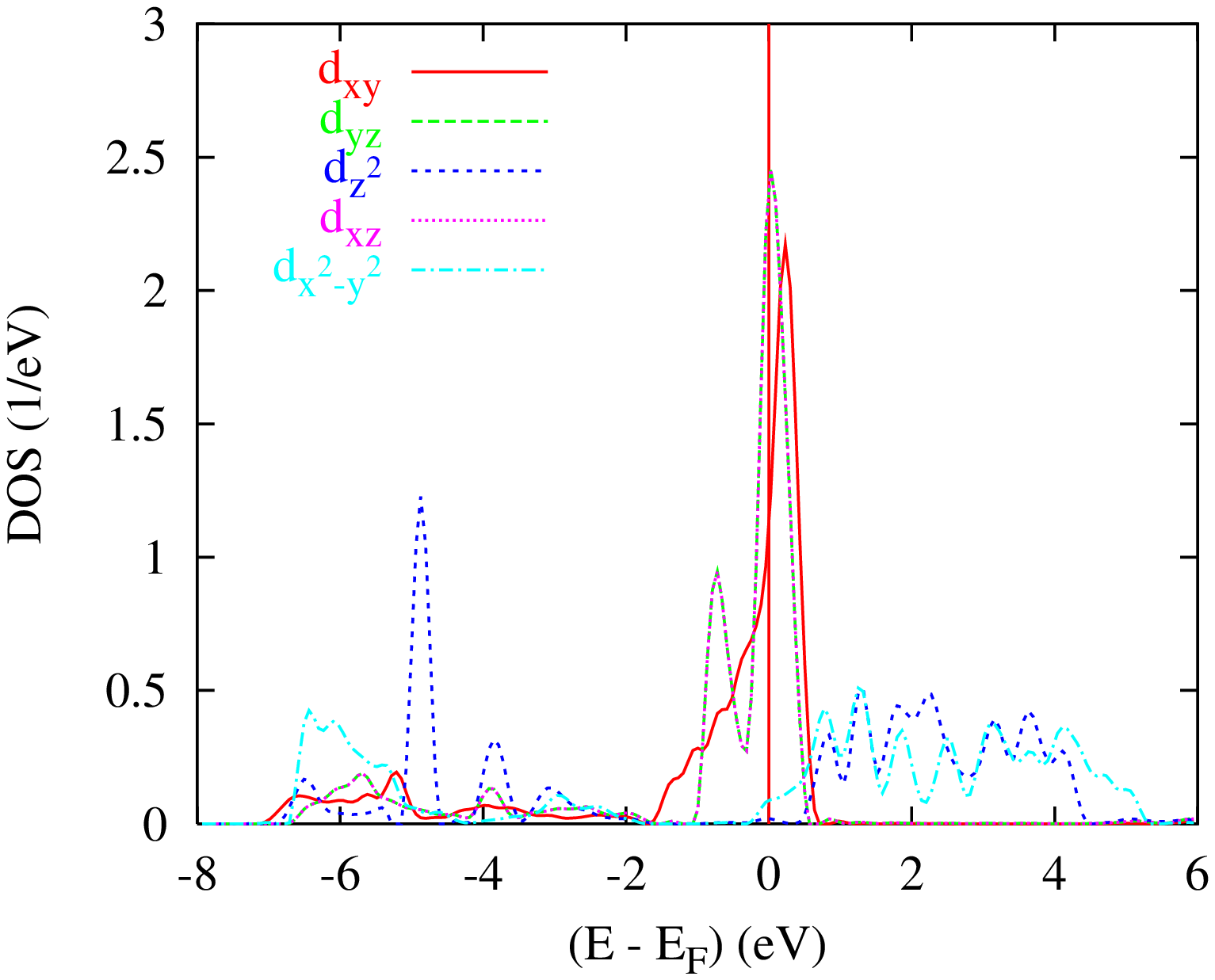}}{a)}
\hspace{0.01\textwidth}
\subfigure{\includegraphics[width=0.42\textwidth]{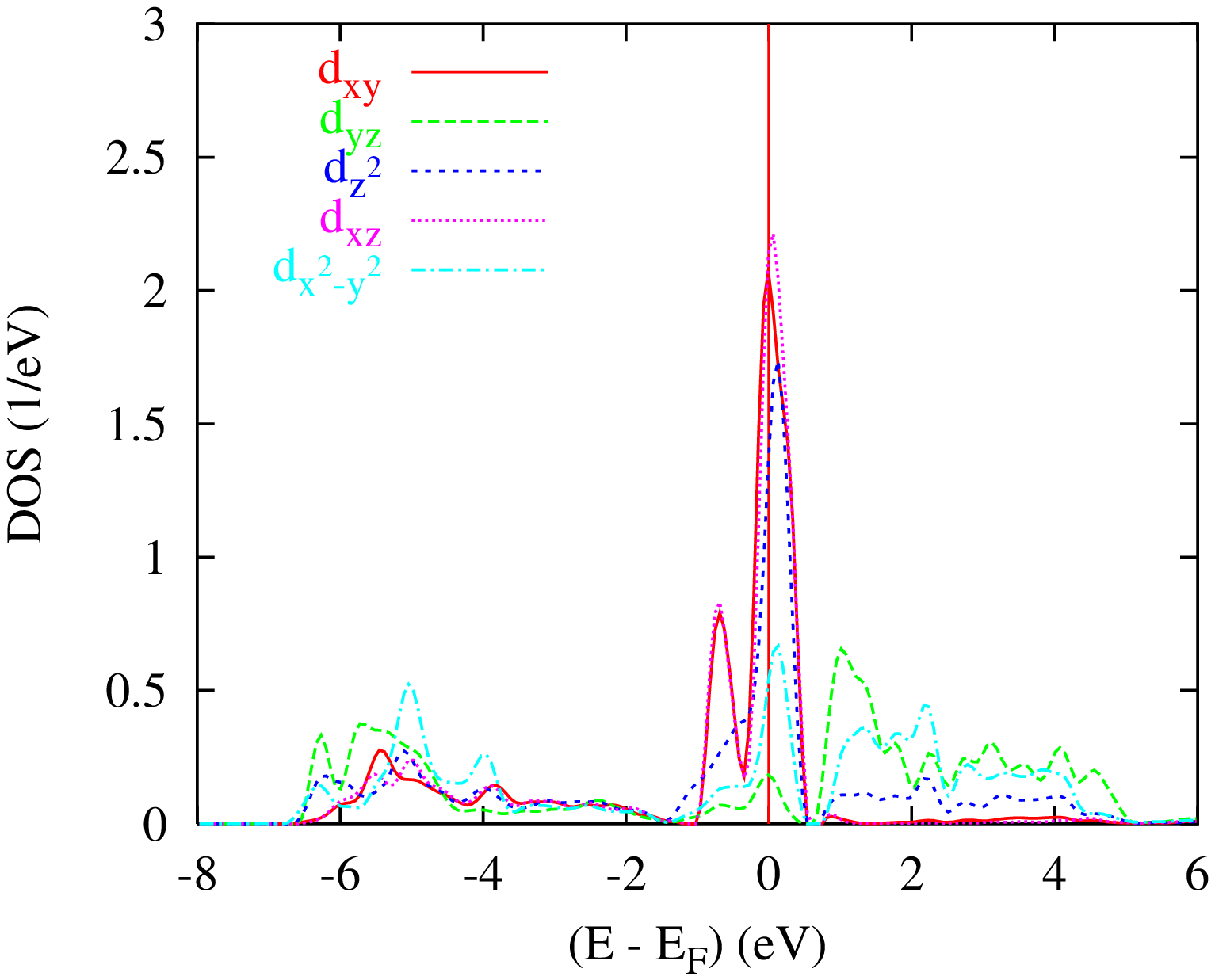}}{b)}
\caption{{\it (color online)}
 Projected DOS plots of the Mn-3d orbitals calculated for the NM states of (a) the minority tetragonal and (b) the majority orthorhombic phases of  Ca$_{3}$Mn$_{2}$O$_{7}$.}
\label{fig:dosCF}
\end{figure}

\begin{figure}[htp]
\centering
\subfigure{\includegraphics[width=0.42\textwidth]{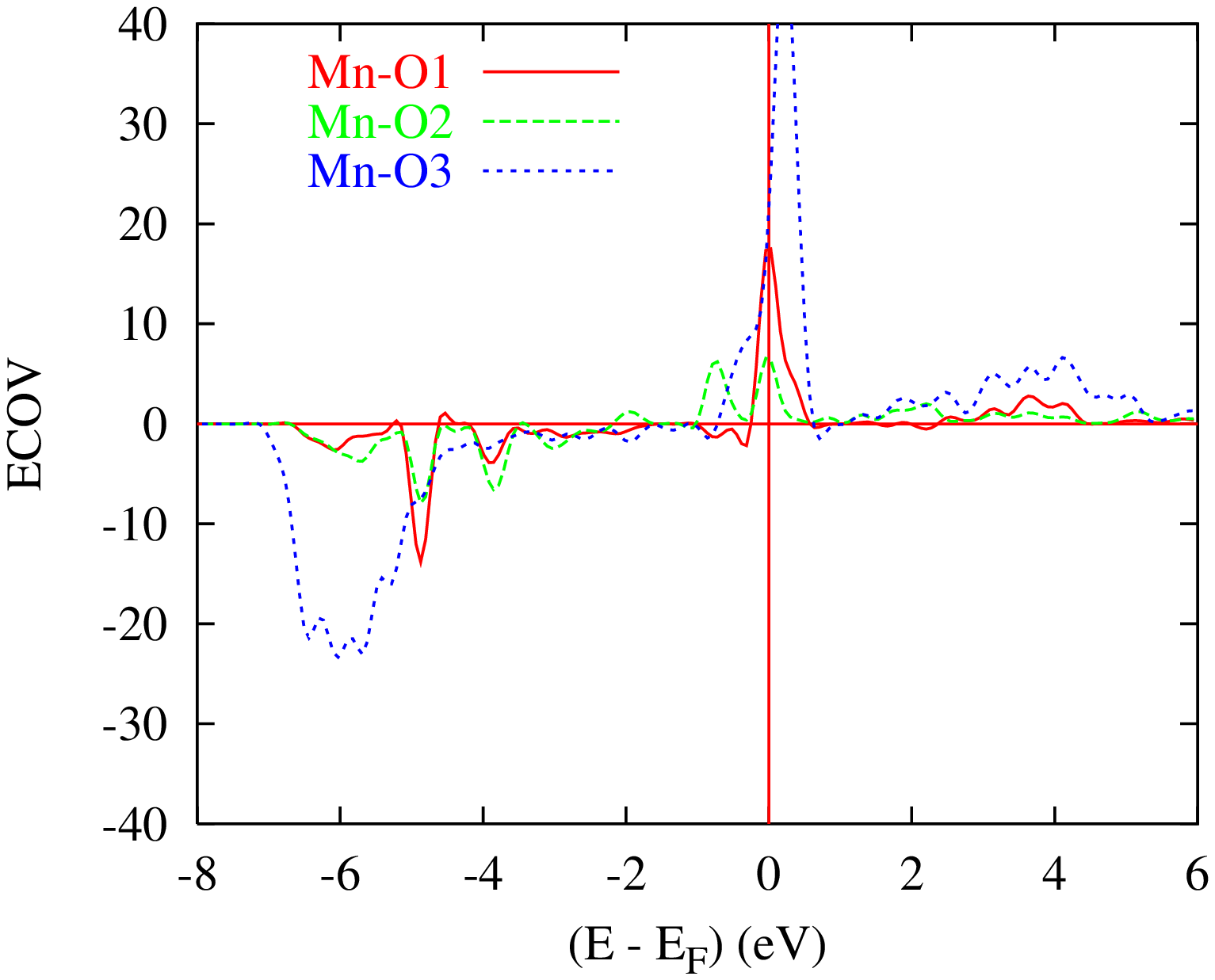}}{a)}
\hspace{0.01\textwidth}
\subfigure{\includegraphics[width=0.42\textwidth]{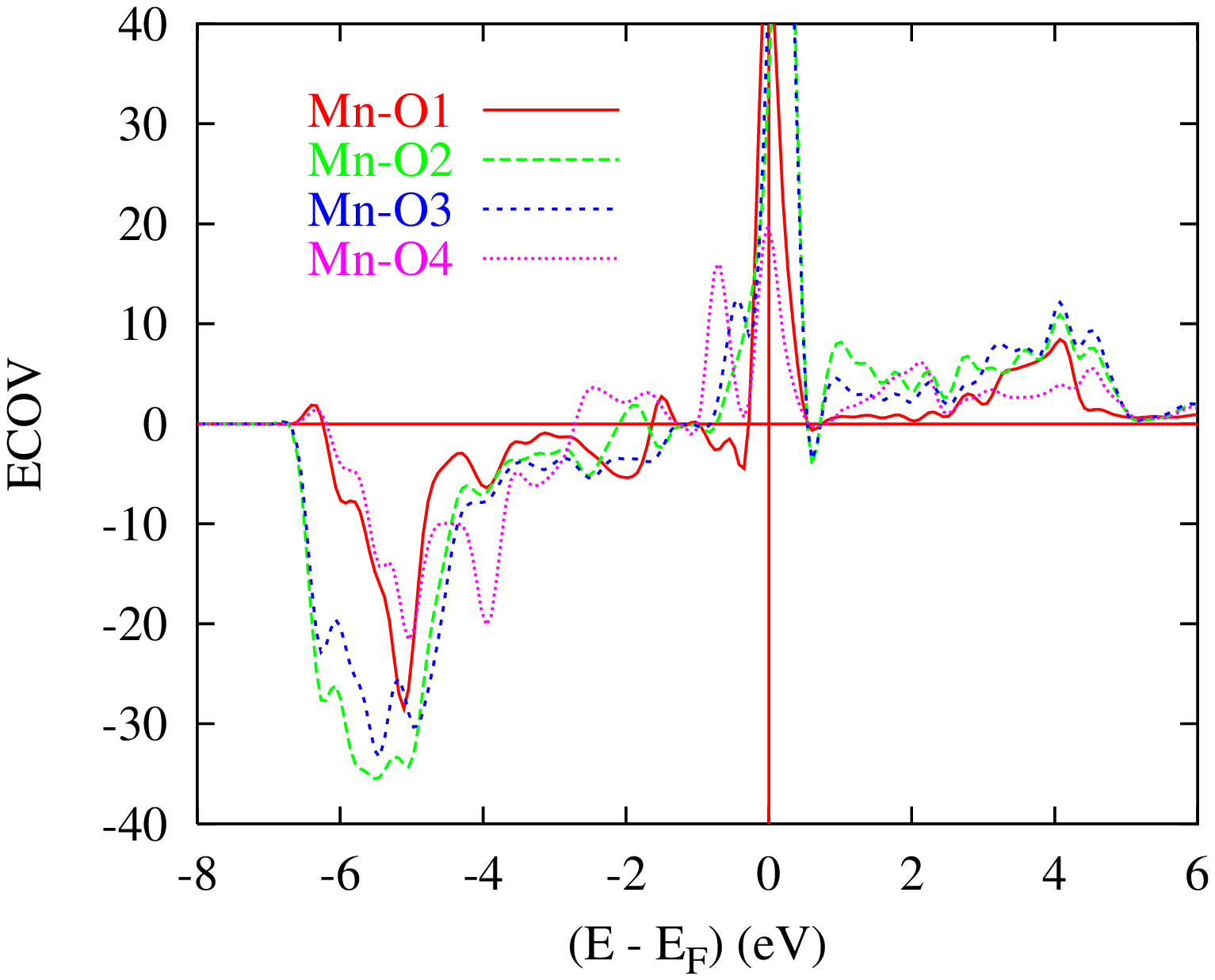}}{b)}
\caption{{\it (color online)}
ECOV plots of the Mn-O bonds calculated for the NM states of (a) the minority tetragonal and (b) the majority orthorhombic phases of Ca$_{3}$Mn$_{2}$O$_{7}$.}
\label{fig:covCF}
\end{figure}
\begin{figure}[htp]
\centering
\includegraphics[width=1.\textwidth]{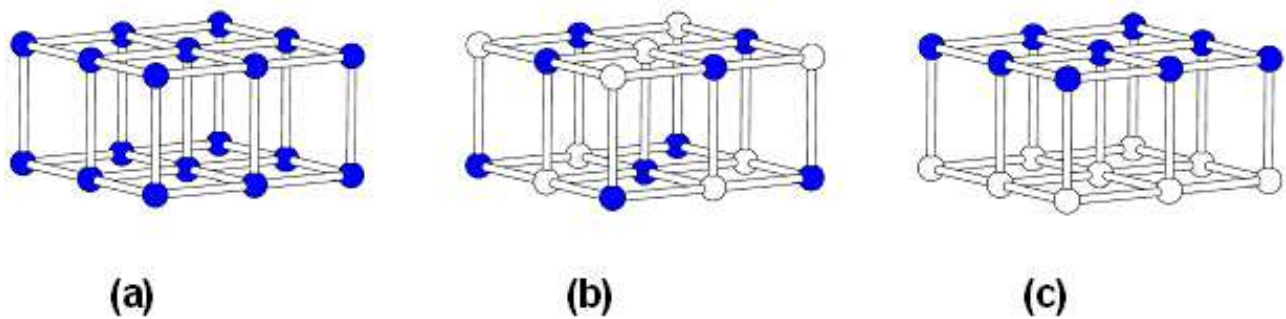}
\caption{{\it (color online)} Schematic representations of the Mn spin arrangements in the (a) FM, (b) the G-type AF, and (c) A-type AF states of Ca$_{3}$Mn$_{2}$O$_{7}$. Blue and white spheres refer to up- and down- spin alignments respectively.}
\label{fig:spin}
\end{figure}

\begin{figure}[htp]
\centering
\subfigure{\includegraphics[width=0.43\textwidth]{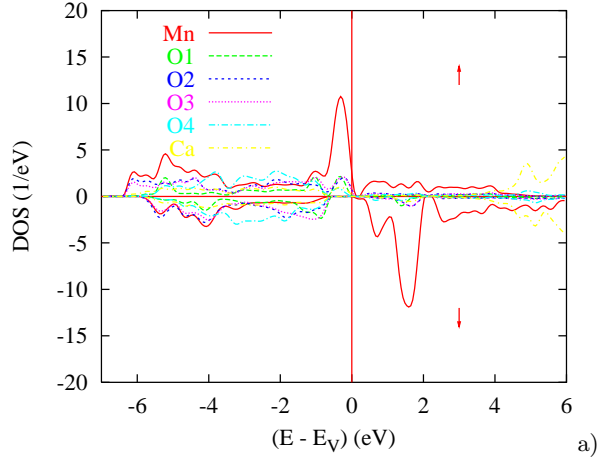}}{a)}
\hspace{0.01\textwidth}
\subfigure{\includegraphics[width=0.43\textwidth]{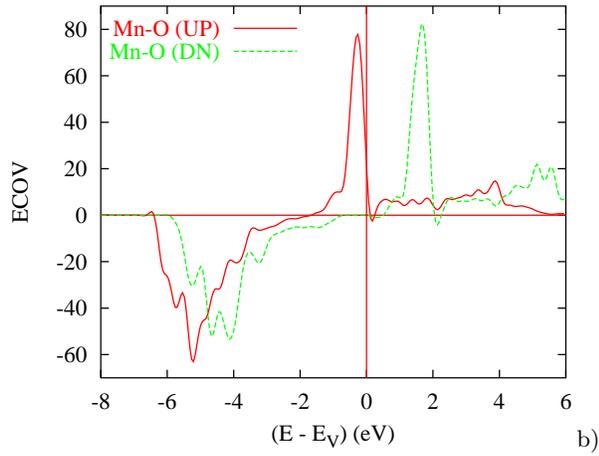}}{b)}
\caption{{\it (color online)}
         FM Ca$_{3}$Mn$_{2}$O$_{7}$. Site and spin projected DOS (a) 
         and chemical bonding (ECOV) for Mn-O interaction regrouping all 
         four oxygen sublattices (b). Energy reference is with respect to the VB top E$_{V}$ due to the closely insulating character.}
\label{fig:bndCF}
\end{figure}

\begin{figure}[htp]
\centering
\subfigure{\includegraphics[width=0.43\textwidth]{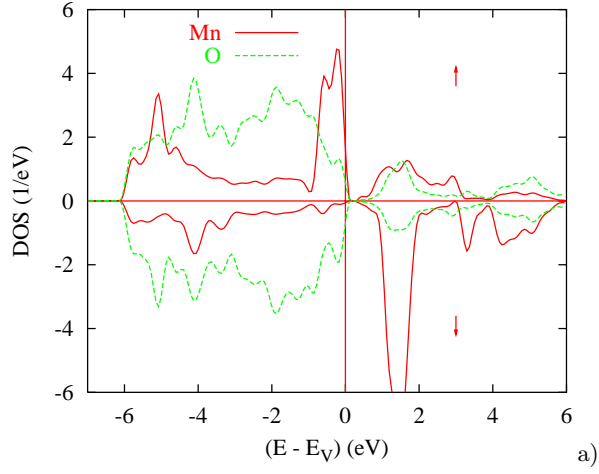}}{a)}
\hspace{0.01\textwidth}
\subfigure{\includegraphics[width=0.45\textwidth]{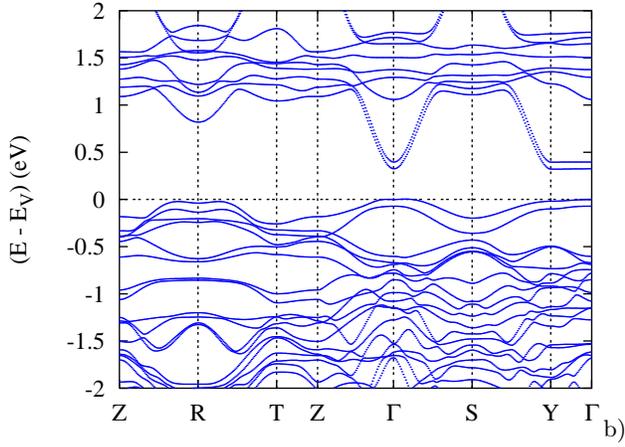}}{b)}
\caption{{\it (color online)}
          AF Ca$_{3}$Mn$_{2}$O$_{7}$ : Site and spin projected DOS (a) and band structure (b). Plots are for one magnetic sublattice. O-PDOS regroup all oxygen sites.}
\label{fig:bndAF}
\end{figure}

\end{document}